\documentstyle[prl,aps,epsbox]{revtex}
\begin{document}
\draft
\preprint{}
\twocolumn[\hsize\textwidth\columnwidth\hsize\csname@twocolumnfalse%
\endcsname 
\title{
Fermion-Boson Duality of
One-dimensional Quantum Particles \\
with Generalized Contact Interaction 
}
\author{
Taksu Cheon${ }^{1,2}$ 
and 
T. Shigehara${ }^{3}$
}
\address{
${ }^1$
Laboratory of Physics, Kochi University of Technology,
Tosa Yamada, Kochi 782-8502, Japan\\
${ }^2$
Theory Division, 
High Energy Accelerator Research Organization (KEK),
Tsukuba, Ibaraki 305-0801, Japan\\
${ }^3$
Department of Information and Computer Sciences,  
Saitama University,
Urawa, Saitama 338-8570, Japan
}
\date{August 28, 1998}
%
\maketitle
\begin{abstract}
For a system of spinless one-dimensional fermions, 
the non-vanishing short-range limit of two-body interaction 
is shown to induce the wave-function discontinuity.
We prove the equivalence of this fermionic system 
and the bosonic particle system with two-body $\delta$-function 
interaction with the reversed role of strong and weak couplings.
%
\end{abstract}
%
]
\narrowtext

%
%
%
The relation between the spin and the exchange statistics
is one of the fundamental properties of particles residing
in four-dimensional Minkowski space.
In lower dimension, however, the relation becomes blurred,
as evidenced in the appearance of anyons in the system 
of two spatial dimension.
In spatial dimension one, the relation lose its meaning
since the spin itself is rather a phenomenological concept
having no ground in the representation theory of Lorentz group.
The discovery of the strict equivalence 
of bosonic sin-Gordon model and 
fermionic massive Thirring model \cite{CO75} suggests that
the exchange statistics also is no absolute concept in one
spatial dimension.  Aside from its aesthetic value, 
this equivalence has practical ramifications in the treatment 
of interacting many-body system in lower dimension. 
There, the relevant aspect is the fact 
that the strong coupling in fermionic model corresponds 
to the weak coupling in the bosonic model and vice versa.

There is indeed a historic precedence 
to the bosonization of fermionic theory 
in a setting of quantum many-body problem.
In the Tomonaga-Luttinger theory 
of one-dimensional fermi liquid \cite{TO50,LU63,ML65,HA81}, 
low energy excitations are describable in terms of
bosonic degrees of freedom. Despite its status as a 
classical standard, the model has several drawbacks.
Firstly, the equivalence is exact only for the
ground state of the system. 
Another problem is its non-applicability to the short-range 
interaction as noted in the original paper by Tomonaga.  
This makes a sharp contrast to the case of bosons 
in one dimension where a simple but rich model of 
particles with two-body $\delta$-interaction
exists \cite{LL63}, whose solvability allows the physical
intuition as well as the thorough thermodynamical analysis.

The purpose of this paper is to formulate a model of fermionic 
many-body system in one-dimension with 
non-vanishing zero-range interaction.
Its analysis reveals that the model can be exactly mapped 
to the same number of bosonic particles interacting 
through $\delta$-interaction with the strength of the coupling
reversed.  This means that we have had a solvable model of
interacting fermions for quite some time without recognizing 
it as such.  It gives a tractable
model of one-dimensional system with non-trivial property of
fermion-boson duality.

%
%
%
We start with a very elementary setting of two identical 
particles with unit mass in one dimension
obeying the fermi statistics.
The wave function of the system has the property
\begin{eqnarray} 
\label{1} 
\Psi _-(x_1,x_2)=-\Psi _-(x_2,x_1) ,
\end{eqnarray}
where $x_1$ and $x_2$ denote 
the coordinates of the particles.
Let us suppose that the two particles are interacting
through a two-body potential $V(x_1-x_2)$.
For now, we place one-body harmonic interaction 
for the technical convenience to bind
the system around the origin.
The Schr{\"o}dinger equation is given by
\begin{eqnarray} 
\label{2} 
\left[ {
\sum_{i=1}^{2}{ 
         \left({ -{1 \over 2}{{d^2} \over {dx_i^2}}
                +{1\over 2}\omega^2 x_i^2}
         \right)}
+V(x_1-x_2)
} \right]\Psi _-(x_1,x_2)
\\ \nonumber
=E\Psi _-(x_1,x_2) .
\end{eqnarray}
With the usual use of the relative and center-of-mass
coordinates $x=x_2-x_1$ and $X=(x_1+x_2)/ 2$,
the system separates into two subsystems as
\begin{eqnarray} 
\label{3} 
\Psi _-(x_1,x_2)=\varphi _-(x)\Phi (X) ,
\end{eqnarray}
where the center-of-mass 
wave function $\Phi (X)$, given by
\begin{eqnarray} 
\label{4}
 \left[ {
    -{1 \over 4}{{d^2} \over {dX^2}}
    +\omega^2 X^2 
        } \right] \Phi (X)
= E^C\Phi (X) 
\end{eqnarray}
is trivial,
and the physics is in 
the relative wave function $\varphi_-(x)$, which satisfies
\begin{eqnarray} 
\label{5} 
\left[ {-{{d^2} \over {dx^2}}+{1\over 4}\omega^2 x^2 +V(x)
       }  \right]\varphi _-(x) 
=E^r\varphi _-(x) .
\end{eqnarray}
The identity of the particles requires $V$ to be symmetric
\begin{eqnarray} 
\label{6} 
V(-x)=V(x).
\end{eqnarray}
The fermionic exchange symmetry, Eq. (\ref{1}), now reads
\begin{eqnarray} 
\label{7}
\varphi _-(-x)=-\varphi _-(x) .
\end{eqnarray}
We consider the case where the potential is short-ranged.  
Namely 
\begin{eqnarray} 
\label{8}
V(x)=0 
\ \  {\rm if} \ \ |x|>a
\end{eqnarray}
for a small positive number $a$.
At the limit $a \to 0$, the self-adjoint 
extension theory dictates 
that any Hermitian potential has to be reduced to the generalized 
pointlike interaction \cite{GK85,SE86,SE86a,AG88}
\begin{eqnarray} 
\label{9} 
V(x)\to \chi (x;\alpha ,\beta ,\gamma ,\delta )
\end{eqnarray}
which admits the discontinuity both of the wave
function and its space-derivative.
The physical understanding of 
this rather counter-intuitive object
$\chi$ is possible through  
its explicit construction in terms of local operator, 
which has recently been 
devised \cite{CS98}.
We have 
\begin{eqnarray} 
\label{10}
& &
\chi (x;\alpha ,\beta ,\gamma ,\delta )
\\ \nonumber
&=&
\mathop {\lim}\limits_{a\to +0}\left[ {u_-
\delta (x+a)+u_0\delta (x)+u_+\delta (x-a)} \right] ,
\end{eqnarray}
where the strengths of $\delta$-functions are given by
\begin{eqnarray} 
\label{11}
u_+(a) &=& -{1 \over {a}}+{{\alpha -1} \over {\delta }} ,
\\ \nonumber
u_-(a) &=& -{1 \over {a}}+{{\gamma -1} \over {\delta }} ,
\\ \nonumber
u_0(a) &=& {{1-\alpha \gamma } \over {\beta a^2}} ,
\end{eqnarray}
in which $\alpha$, $\beta$, $\gamma$ and $\delta$ 
are arbitrary real numbers with the constraint
\begin{eqnarray} 
\label{12} 
\alpha \gamma -\beta \delta =1 .
\end{eqnarray}
The effect of the $\chi(x)$ on the wave function 
can be expressed as
\begin{eqnarray} 
\label{13}
\varphi _-'(0_+)+\alpha \varphi _-'(0_-)
=-\beta \varphi _-(0_-)
\\ \nonumber
\varphi _-(0_+)+\gamma \varphi _-(0_-)
=-\delta \varphi _-'(0_-) .
\end{eqnarray}
From the antisymmetry of $\varphi(x)$, Eq. (\ref{7}), 
one has $\varphi'(-x)$ $=\varphi'(x)$.  This
implies $\alpha = -1$ and $\beta = 0$,
which, in combination with  Eq. (\ref{12}), 
results in $\gamma=-1$.
We therefore have
\begin{eqnarray} 
\label{14} 
V(x)\varphi _-(x)\to \varepsilon (x;c)\varphi _-(x)
\ \ \ (a\to 0)
\end{eqnarray}
where $\varepsilon(x;c)$ is defined by
\begin{eqnarray} 
\label{15} 
\varepsilon (x;c)\equiv \chi (x;-1,0,-1,-4c) .
\end{eqnarray}
This interaction induces the 
discontinuity in the {\it wave function itself} whose
amount is specified by the real number number $c$ through
\begin{eqnarray} 
\label{16}
\varphi _-(0_+)=-\varphi _-(0_-)
= 2c\varphi _-'(0_+)= 2c\varphi _-'(0_-) .
\end{eqnarray}
In place of Eq. (\ref{14}), one can also 
have $\chi(x; 1, \beta, 1, 0)$
as a legitimate zero-range limit. 
However, its effect on $\varphi_-(x)$ is 
exactly the same as $\varepsilon(x; 1/\beta)$. 

An explicit construction of $\varepsilon(x; c)$ is obtained
from Eq. (\ref{11}) as
\begin{eqnarray} 
\label{17}
& &
\varepsilon (x;c)\varphi _-(x)
\\ \nonumber
&=&
\mathop {\lim}\limits_{a\to +0}
 \left( {{1 \over {2c}}-{1 \over {a}}} \right)
 \left\{ {\delta (x+a)+\delta (x-a)} \right\}\varphi _-(x) .
\end{eqnarray}
This should be contrasted to the ``usual'' zero-range limit,
Dirac's delta function
\begin{eqnarray} 
\label{18} 
\delta (x;v)\equiv \chi (x;-1,-v,-1,0) 
=v\delta (x) ,
\end{eqnarray}
which has no effect on the antisymmetric wave function;
\begin{eqnarray} 
\label{19} 
\delta (x;v)\varphi _-(x)=0 .
\end{eqnarray}
Thus the non-vanishing zero-range
limit of the system is described by
\begin{eqnarray} 
\label{20}
\left[ {-{{d^2} \over {dx^2}}
        +{1\over 4}\omega^2 x^2 + \varepsilon (x;c)
} \right]
\varphi _-(x)
=E^r\varphi _-(x) .
\end{eqnarray}
%

%
\begin{figure}
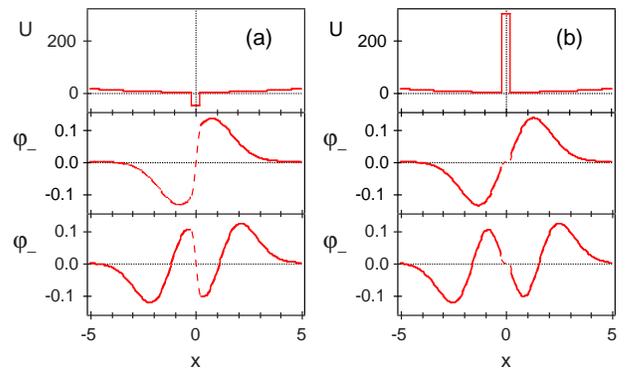

\label{fig1}
\center\psbox[hscale=0.450,vscale=0.450]{dufig1.epsf}
\caption{
Examples of fermionic relative wave functions.
The harmonic oscillator parameter $\omega=2$ sets the
scale.  The interaction $V$ is chosen to be a square well
with the range $a = 0.2$.
(a) The case for attractive interaction of depth $-50$.
(b) The case of repulsive interaction of height $300$.
In both cases, top figures are the profile 
of the interaction $U$ $=\omega^2x^2/4$ $+V$.
  The middle and the bottom
are the lowest and the second lowest energy eigenfunctions
}
\end{figure}
Intuitive meaning of the admissibility of the 
discontinuity-inducing 
interaction $\varepsilon(x;c)$ should become clear 
by inspecting Fig. 1, where we depict the antisymmetric
wave functions subjected to a symmetric potential of small
but finite range.  Fig. 1(a) is an example of the attractive
potential, and Fig. 1(b) a repulsive one.
The procedure, Eqs. (\ref{16})-(\ref{17}) is
a non-trivial but sensible zero-range limit
that keeps the non-vanishing effect of the potential
through a rather unfamiliar concept of 
wave-function discontinuity.

We now perform a transformation
\begin{eqnarray} 
\label{21} 
\varphi _+(x)
=\left[ {\theta (x)-\theta (-x)} \right]\varphi _-(x) ,
\end{eqnarray}
where $\theta (x)$ is the step function 
$\theta (x)=1$ when $x>0$ and $\theta (x)=0$ when $x<0$.
The connection condition Eq. (\ref{16}) is rewritten as
\begin{eqnarray} 
\label{22} 
\varphi _+'(0_+)=-\varphi _+'(0_-)
={1 \over {2c}}\varphi _+(0_+)
={1 \over {2c}}\varphi _+(0_-) ,
\end{eqnarray}
which means that $\varphi _+(x)$  satisfies Eq.(\ref{13}) 
with $\alpha =\gamma =-1$, $\delta =0$ and $\beta=-1/c$.  
In other words, $\varphi_+(x)$
is a solution of the Schr{\" o}dinger equation 
\begin{eqnarray} 
\label{23}
\left[ {-{{d^2} \over {dx^2}} 
        +{1\over 4}\omega^2 x^2 + \delta (x;v)
       } \right]\varphi _+(x)
=E^r\varphi _+(x) ,
\end{eqnarray}
if the coupling constants $v$ and $c$ are related by
\begin{eqnarray} 
\label{24}
v={1 \over {c}} .
\end{eqnarray}
By construction, one has
\begin{eqnarray} 
\label{25} 
\varphi _+(-x)=\varphi _+(x) .
\end{eqnarray}
In terms of the full two-particle wave function
\begin{eqnarray} 
\label{26} 
\Psi _+(x_1,x_2)=\varphi _+(r)\Phi (x) ,
\end{eqnarray}
this signifies the bosonic exchange symmetry
\begin{eqnarray} 
\label{27} 
\Psi _+(x_1,x_2)=\Psi _+(x_2,x_1) .
\end{eqnarray}
Therefore, two-fermion system with $\varepsilon$-interaction
is equivalent to two-boson system with $\delta$-interaction,
and the strong coupling in one side corresponds to 
the weak coupling in the other.  
We emphasize that $\delta$ and $\varepsilon$ functions
are the {\it only} non-vanishing limits of any interaction 
that acts on bosonic and fermionic wave functions
respectively.
Note the parallel relation to Eq. (\ref{19})
for $\varphi _+(x)$;
\begin{eqnarray} 
\label{28} 
\varepsilon (x;c)\varphi _+(x)=0 .
\end{eqnarray}
Note also that the couplings $v$ and $c$ can be both 
positive and negative.  In the latter case, the equivalence
extends to the negative-energy bound states that exist in
both fermi and bose systems. 
%
\begin{figure}
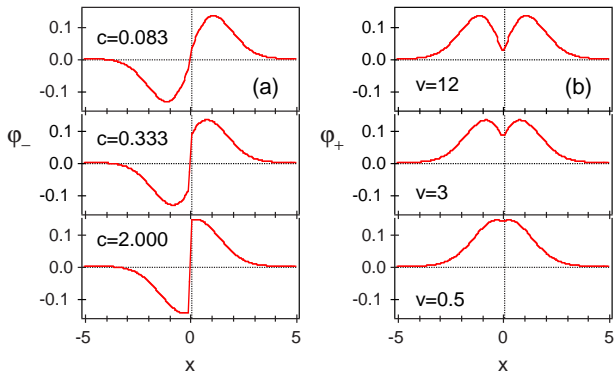

\label{fig2}
\center\psbox[hscale=0.450,vscale=0.450]{dufig2.epsf}
\caption{
The fermionic (a) and bosonic (b) relative wave functions 
with three values of coupling parameter.  At each raw,
(a) and (b) are related by the transformation, Eq. (21).
The $\varepsilon$-interactions for case (a)
are constructed from  Eq. (17) with $a \to 0$ limit 
replaced by a small number $a= 0.05$.
}
\end{figure}

It is instructive to look at the wave functions to see the 
actual workings of the boson-fermion duality  
with some numerical examples.  
We show in Fig. 2(a), the lowest energy fermionic
eigenstates of Eq. (\ref{20}) with several values 
of coupling strengths.  In Fig. 2(b), the corresponding
bosonic eigenstates of Eq. (\ref{23}) are displayed. 
In the calculation, the $\varepsilon$-
interaction, Eq. (\ref{17}), is evaluated with small 
but finite value of $a$ in place of $a \to 0$ limit.
These figures show that the rigorous
results at the mathematical limit $a \to 0$ does
have real relevance to more realistic problem
with finite-range interactions.

%
%
%
It is straightforward to extend the above arguments
to the system of $N$ one-dimensional particles.
Let us write the wave function of the system 
for the particular ordering of 
the set of $N$ coordinates $(x_1, x_2, ... , x_N)$, say
$x_1$ $>$ $x_2$ $>$ $\cdot\cdot\cdot$ $>$ $x_N$, 
as $\Psi_1$;
\begin{eqnarray} 
\label{29} 
\Psi _1 \equiv \Psi (x_1,...,x_N)
 \theta(x_1-x_2) \cdot\cdot\cdot  \theta(x_{N-1}-x_N)  .
\end{eqnarray}
We define the permutation $P$ of $N$ numbers
\begin{eqnarray}
\label{30} 
P : (1,2,...,N) \to (P_1, P_2, ..., P_N). 
\end{eqnarray}
Suppose $(-1)^P$ represents the parity of the permutation $P$.
The wave functions $\Psi_\pm$ defined by
\begin{eqnarray} 
\label{31} 
\Psi _\pm(x_1,...,x_N)
= {1\over\sqrt{N!}} \sum_P (\pm 1)^P\Psi_1 (x_{P_1},...,x_{P_N})
\end{eqnarray}
have the exchange symmetry
\begin{eqnarray} 
\label{32} 
\Psi _\pm(...,x_i,...,x_j,...) 
= \pm \Psi _\pm(...,x_j,...,x_i,...) .
\end{eqnarray}
Namely, $\Psi_+$  and $\Psi_-$ represent 
the systems of $N$ bosons and $N$ fermions, respectively.
It is easy to see that following two equations are equivalent;
\begin{eqnarray} 
\label{33} 
\left. {\Psi _-} \right|_{x_i=x_{j+}}
&=& -\left. {\Psi _-} \right|_{x_i=x_{j-}}
\\ \nonumber
&=&
c\left.{ 
        \left({ 
                {\partial  \over {\partial x_i}}
               -{\partial  \over {\partial x_j}} 
              }\right)
        \Psi _-
        }\right|_{x_i=x_{j+}}
\\ \nonumber
&=&
c\left.{ 
        \left({ 
                {\partial  \over {\partial x_i}}
               -{\partial  \over {\partial x_j}} 
              }\right)
        \Psi _-
      }\right|_{x_i=x_{j-}} ,
\end{eqnarray}
\begin{eqnarray} 
\label{34} 
& &
\ \ \ 
\left.{ 
      \left({ 
              {\partial  \over {\partial x_i}}
             -{\partial  \over {\partial x_j}} 
            }\right)
      \Psi _+
      }\right|_{x_i=x_{j+}}
\\ \nonumber
&=&
-\left.{ 
      \left({ 
              {\partial  \over {\partial x_i}}
             -{\partial  \over {\partial x_j}} 
            }\right)
      \Psi _+
      }\right|_{x_i=x_{j-}}
\\ \nonumber
&=&
{1\over {c}} \left. {\Psi _+} \right|_{x_i=x_{j+}}
= {1\over {c}} \left. {\Psi _+} \right|_{x_i=x_{j-}} .
\end{eqnarray}
Therefore, $\varepsilon(x_i-x_j;c)$ acting on $\Psi_-$ 
and $\delta(x_i-x_j;1/c)$ acting on $\Psi_+$
are two different representations of the same effect.  
We have the equivalence of two equations,
\begin{eqnarray} 
\label{35} 
\left[ {
\sum_{i} 
        {\left({ -{1 \over 2}{{d^2} \over {dx_i^2}}
                +{1\over 2}\omega^2 x_i^2}
         \right)}
+\sum_{i>j}\varepsilon(x_i-x_j; c)
} \right]
\\ \nonumber
\times
\Psi _-(x_1,...,x_N)=E\Psi _-(x_1,...,x_N)
\end{eqnarray}
and
\begin{eqnarray} 
\label{36} 
\left[ {
\sum_{i} 
        {\left({ -{1 \over 2}{{d^2} \over {dx_i^2}}
                +{1\over 2}\omega^2 x_i^2}
         \right)}
+\sum_{i>j}\delta(x_i-x_j; {1\over{c}})
} \right]
\\ \nonumber
\times
\Psi _+(x_1,...,x_N)=E\Psi _+(x_1,...,x_N),
\end{eqnarray}
that can be mapped into each other.
 
%
%
%
The confining harmonic potential is an artifact to
supply the basis functions, 
which sometimes causes a nuisance.
Alternatively, one sets $\omega = 0$ and imposes
the cyclic boundary condition
\begin{eqnarray} 
\label{37} 
\Psi ( \cdot\cdot, x_i+L,\cdot\cdot)
=\Psi ( \cdot\cdot, x_i,\cdot\cdot) 
\ \ \ {\rm for} \ \ \ i=1,..,N.
\end{eqnarray}
There is a subtle complication with this 
prescription \cite{LL63,YA67}, which we analyze
in the followings.
Let us suppose, for a moment, that we have a set of $x_i$ 
all within the range of length $L$, 
say ${L\over 2}$  $> x_i$ $>-{L\over 2}$ with the ordering
$x_1 $ $>x_2$ $>\cdots$ $>x_N$.
By definition, one has
$\Psi_\pm(x_1,...,x_N)$ $=\Psi_1(x_1,...,x_N)$.
With the replacement $x_N \to x_N+L$ one has
$\Psi_\pm(x_1,...,x_N+L)$ 
$=(\pm1)^{N-1}\Psi_1(x_N+L,x_1,...,x_{N-1})$.
This can be rewritten as a relation between $\Psi_+$
and $\Psi_-$ in the form
$\Psi_-(x_1,...,x_N+L)$ 
$=(-1)^{N-1}\Psi_+(x_1,...,x_N+L)$.
Thus it is not always appropriate to impose the 
cyclic boundary both for $\Psi_+$ and  $\Psi_-$.
A consistent description of the boundary is achieved by 
replacing the strict periodic condition Eq. (\ref{37}) 
with a relaxed version
\begin{eqnarray} 
\label{38} 
&\Psi_\pm& ( \cdot\cdot, x_i+L,\cdot\cdot)
= e^{i\lambda_\pm}\Psi_\pm ( \cdot\cdot, x_i,\cdot\cdot) 
\ \ {\rm for} \ \ i=1,..,N
\\ \nonumber
&{\rm with}&
\\ \nonumber
&\lambda_-& =\lambda_+ + (N-1)\pi.
\end{eqnarray}
Then, for $\omega = 0$, the fermionic problem Eq. (\ref{35}) 
is equivalent to the bosonic problem Eq. (\ref{36}).
Specifically, the usual choice $\lambda_+=0$ gives the
periodic boundary for $\Psi_+$ and antiperiodic boundary
for $\Psi_-$.

The representation of our model in the second-quantized
form should be very useful, since it is in that form
that the bosonization of fermion systems is discussed
with formidable mathematical machinery \cite{IA87,KO97}. 
Also, it could lead to a
new type of field theoretical models.  
A technical block on its way is the non-perturbative nature
of the $\varepsilon$-interaction,
which does not allow meaningful calculations of
its matrix elements, 
at least in naive, straightforward approach.
%
%
%
%

Since the complete solution based on the Bethe ansatz
for the bosonic problem 
Eq. (\ref{36}) with $\omega=0$ exists \cite{LL63},
we now have a model of solvable 
fermion $N$-body problem with non-trivial characteristics.  
It is of particular interest to investigate 
the thermodynamic properties of this system in detail.
 
It would be worthwhile to place our approach in
the context of other solvable many-body models 
in one dimension \cite{OP83}.  In particular,
the study of its relation (or contrast) to the
model with a long-range interaction, namely the 
Calogero-Sutherland model \cite{SU71,CA71,HA96}, 
appears to be a promising subject.    
It should be also interesting to look at the
fermion-boson relations 
in {\it other} dimensions.  This is especially true
in light of 
a recent work on the equivalence between free fermions
and free bosons in dimension two \cite{LE97}.
Finally, we would like to call readers attention to
rather unexplored 
potential roles of the generalized contact 
interactions in other contexts than discussed here.
Those include such diverse subjects as the 
semiconductor heterojuctions \cite{BB95} and the
controversy over the one-dimensional
$1/|x|$ potential \cite{LO59,GC97}.

\vspace*{5mm}
We are very grateful to Prof. T. Tsutsui and 
Dr. T. F\"ul\"op for enlightening discussions and
valuable comments.
Thanks are also due to Prof. Y. Okada, 
Prof. K. Takayanagi and 
Prof. T. Kawai for useful communications.
This work has been supported in part by the Grant-in-Aid
(No. 10640396) by the Japanese Ministry of Education.

\begin{references}
%
\bibitem{CO75}
S. Coleman, Phys. Rev. {\bf D11} (1975) 2088.
%
\bibitem{TO50}
S. Tomonaga, Progr. Theor. Phys. {\bf 5} (1950) 544.
%
\bibitem{LU63}
J. M. Luttinger, J. Math. Phys. {\bf 4} (1963) 1154.
%
\bibitem{ML65}
D.C. Mattis and E. Lieb, J. Math. Phys. {\bf 6} (1965) 304.
%
\bibitem{HA81}
F.D.M. Haldane, J. Phys. {\bf C14} (1981) 2585.
%
\bibitem{LL63}
E. Lieb and W. Linger, Phys. Rev. {\bf 130} (1963) 1605; 
{\it ibid.} 1616. 
%
\bibitem{GK85}
F. Gesztesy an W. Kirsch, J. Reine Angew. Math. 
{\bf 362} (1985) 28. 
%
\bibitem{SE86}
P. {\v S}eba, 
Czech. J. Phys. {\bf B36} (1986) 667. 
%
\bibitem{SE86a}
P. {\v S}eba, 
Rep. Math. Phys. {\bf 24} (1986) 111. 
%
\bibitem{AG88}
S. Albeverio, F. Gesztesy, R. H\o egh-Krohn and H. Holden, 
{\em Solvable Models in Quantum Mechanics} 
(Springer, Heidelberg, 1988).
%
\bibitem{CS98}
T. Cheon and T. Shigehara, 
Phys. Lett. {\bf A243} (1998) 111.
%
\bibitem{YA67}
C.N. Yang, 
Phys. Rev. Lett. {\bf 19} (1967) 1312. 
%
\bibitem{IA87}
F. Iachello and A. Arima, 
{\em The Interacting Boson Model}
(Cambridge Univ. Press, Cambridge, 1987). 
%
\bibitem{KO97}
P. Kopietz, 
{\em Bosonization of Interacting Fermions in 
Arbitrary Dimensions}  
(Springer, Heidelberg, 1997). 
%
\bibitem{OP83}
M.A. Olshanetsky and A.M. Perelomov, 
Phys. Rep. {\bf 94} (1983) 313. 
%
\bibitem{SU71}
B. Sutherland, J. Math. Phys. {\bf 12} (1971) 246; {\it ibid.} 251.
%
\bibitem{CA71}
F. Calogero, J. Math. Phys. {\bf 12} (1971) 419.
%
\bibitem{HA96}
Z.N.C. Ha, {\em Quantum Many-Body Systems in One Dimension} 
(World Scientific, Singapore, 1996).
%
\bibitem{LE97}
M.H. Lee,
Phys. Rev. {\bf E55} (1997) 1518.
%
\bibitem{BB95}
R. Balian, D. Bessis and G.A. Mezincescu,
Phys. Rev. {\bf B51} (1995) 17624.
%
\bibitem{LO59}
R. Loudon, Amer. J. Phys. {\bf 27} (1959) 649.
%
\bibitem{GC97}
A.N. Gordeyev and S.C. Chhajlany, 
J. Phys. {\bf A30} (1997) 6893.
%
\end{references}
\end{document}